# Multipath optical recombination of intervalley dark excitons and trions in monolayer WSe₂

Erfu Liu[1], Jeremiah van Baren[1], Ching-Tarng Liang[2], Takashi Taniguchi[3], Kenji Watanabe[4], Nathaniel M. Gabor[1,5], Yia-Chung Chang[2], Chun Hung Lui[1*]

[1] Department of Physics and Astronomy, University of California, Riverside, CA 92521, USA.

[2] Research Center for Applied Sciences, Academia Sinica, Taipei 11529, Taiwan.

[3] International Center for Materials Nanoarchitectonics (WPI-MANA), National Institute for Materials Science (NIMS), 1-1 Namiki Tsukuba, Ibaraki 305-0044, Japan.

[4] National Institute for Materials Science (NIMS), 1-1 Namiki Tsukuba, Ibaraki 305-0044, Japan.

[5] Canadian Institute for Advanced Research, MaRS Centre West Tower, 661 University Avenue, Toronto, Ontario ON M5G 1M1, Canada.

[*]Corresponding author. Email: joshua.lui@ucr.edu

Abstract:

Excitons and trions (or exciton-polarons) in transition metal dichalcogenides (TMDs) are known to decay predominantly through intravalley transitions. Electron-hole recombination across different valleys can also play a significant role in the excitonic dynamics, but intervalley transitions are rarely observed in monolayer TMDs, because they violate the conservation of momentum. Here we reveal the intervalley recombination of dark excitons and trions through more than one path in monolayer WSe₂. We observe the intervalley dark excitons, which can recombine by the assistance of defect scattering or chiral-phonon emission. We also reveal that a trion can decay in two distinct paths – through intravalley or intervalley electron-hole recombination – into two different final valley states. Although these two paths are energy degenerate, we can distinguish them by lifting the valley degeneracy under a magnetic field. In addition, the intra- and inter-valley trion transitions are coupled to zone-center and zone-corner chiral phonons, respectively, to produce distinct phonon replicas. The observed multipath optical decays of dark excitons and trions provide much insight into the internal quantum structure of trions and the complex excitonic interactions with defects and chiral phonons in monolayer valley semiconductors.

Monolayer transition metal dichalcogenides (TMDs), such as $MoS_2$ and $WSe_2$, exhibit two time-reversal electronic valleys (K, K') with direct band gap in the Brillouin zone [1-12] [Fig. 1(a)]. The electrons and holes in these valleys can form tightly bound excitons that govern the materials' optical properties [13, 14]. Thus far, the related exciton research has focused predominantly on intravalley excitons (consisting of electrons and holes in the same valley) because they are coupled to light. In contrast, intervalley excitons (consisting of electrons and holes in opposite valleys) have not been observed in monolayer TMDs, because their optical transitions are forbidden by momentum mismatch. In principle, intervalley electron-hole recombination can be activated by scattering with defects or phonons [15-22] [Figs. 1(b)-1(d)]. But experimental studies of such scattering-assisted intervalley transitions are still lacking.

In addition, in the presence of a Fermi sea, excitons can couple to the Fermi sea to form trions or exciton-polarons (hereafter, we denote both broadly as "trions") [23-26]. In a generic trion state (e.g., with an electron and two holes), the trion is supposed to decay through two paths – with the electron recombining with either hole – into two different final hole states. But prior research could not distinguish these two paths, though both are essential components of the trion dynamics. In monolayer TMDs, the controllable valleys can help resolve the trion decay paths. A representative trion in monolayer TMDs consists of an electron-hole pair in one valley and a hole in the other valley. The trion can decay through two paths – intravalley or intervalley electron-hole recombination – into a hole in different valleys. By controlling the valley degree of freedom, we might resolve these two paths. So far, only the intravalley trion decay has been revealed, whereas the intervalley trion decay is hidden due to its weak optical signal and energy degeneracy with the intravalley decay. It is therefore of fundamental interest to explore the intervalley trion transition to better understand the trion quantum structure.

In this Letter, we observe intervalley recombination of excitons and trions through multiple (more than one) quantum paths in monolayer $WSe_2$. We find that the intervalley excitons can emit helical luminescence through intervalley transitions mediated by defect scattering or chiral-phonon emission. We also reveal that a trion can decay in two distinct paths – through either intravalley or intervalley electron-hole recombination – into two different final valley states. Although these two decay processes are energy degenerate, we can lift the degeneracy and resolve their distinct emission under a magnetic field. In addition, the intravalley (intervalley) trion transitions can couple selectively with zone-center (zone-corner) chiral phonons to produce distinct replica lines. Such phonon-assisted emission provides extra trion decay paths. Furthermore, by measuring the Zeeman splitting between intra- and inter-valley trion transitions, we can, for the first time, extract the effective mass of the lower conduction band in monolayer $WSe_2$.

Our observations are made possible through ultraclean monolayer $WSe_2$ devices with the $BN/WSe_2/BN$/graphite van der Waals heterostructure, where the encapsulation by hexagonal boron nitride (BN) and the usage of graphite contacts significantly enhance the device performance [27]. Figure 2(a) displays a gate-dependent PL map of monolayer WSe$_2$ with no magnetic field [27]. The PL map features the bright $A$ exciton and trions ($A^0$, $A^+$, $A_1^-$, and $A_2^-$), dark exciton and trions ($D^0$, $D^+$, and $D^-$) [29, 38, 53-59], and the zone-center chiral-phonon replicas of the dark states ($D_p^0$, $D_p^+$, and $D_p^-$) with ~21.4 meV redshift energy [28, 60]. These excitonic features, all from *intravalley* optical transitions, have been studied extensively in the literature. Here the exceptional quality of our device allows us to reveal a set of hitherto unreported PL peaks labeled as $I^0$, $I_p^0$, $I_p^+$, and $I_p^-$ in Fig. 2. Each of these peaks exhibits nearly linear power dependence, which rules out the association with biexcitonic states [27, 61-64]. We have compelling evidence that these observed peaks arise from distinct *intervalley* electron-hole recombination processes of dark excitons and trions.

We attribute the $I^0$ peak at charge neutrality to the intervalley dark excitons based on two observations [Figs. 1(a) and 1(b)]. First, the $I^0$ feature at 1.702 eV is 10 meV higher than the intravalley dark exciton ($D^0$) at 1.692 eV [Figs. 2(a)-(c), Fig. S7 [27]]. The binding energy of an exciton is controlled by the electron-hole Coulomb interaction and exchange interaction [2, 65]. The $D^0$ dark exciton with antiparallel electron spins has no exchange interaction. But the intervalley exciton with parallel spins has finite exchange interaction, which partially counteracts the direct Coulomb interaction and reduces the binding energy [19]. We have carried out first-principles calculations, which show that the binding energy of intervalley exciton is ~10 meV smaller than that of intravalley dark exciton due to the exchange interaction (Sec. 9 of the Supplemental Material [27]). This accounts for the 10 meV blueshift of $I^0$ from $D^0$.

Second, the measured $I^0$ $g$ factor (12.4) matches the predicted $g$ factor of intervalley exciton (~13). Figure 3 shows the Zeeman shift and $g$ factor of different emission lines with right- and left-handed circular polarization under magnetic field $B = 0 – 17$ T. The $g$ factor is defined by $\Delta E = g\mu_B B$, where $\Delta E$ is the energy difference between left- and right-handed emission and $\mu_B$ is the Bohr magneton. The total Zeeman shift includes components from the spin, atomic orbits, and Berry curvature [30-34] (Figs. S8 and S9 [27]). For intravalley excitons, the Berry-curvature contributions from the conduction and valence bands cancel each other; this leads to a predicted $g$ factor of ~4 for $A^0$ exciton and ~8.4 for $D^0$ exciton, roughly consistent with our data [Figs. 3(c) and 3(d)] [29]. But for intervalley excitons, the Berry-curvature contributions from the conduction and valence bands add up. This leads to a larger predicted $g$ factor of ~13 [27]. The value closely matches the observed $I^0$ $g$ factor (12.4) [Figs. 3(c) and 3(d)].

The consistent $I^0 - D^0$ energy separation and $I^0$ $g$ factor strongly support that the $I^0$ peak comes from the intervalley exciton. The momentum-forbidden (but spin-allowed) intervalley transitions can be mediated by defect scattering [18]. Carrier injection can effectively fill the defect states and screen the defect potential. Consistently, we observe $I^0$ sharply at the charge neutrality point, in a carrier-density range much narrower than those of intravalley excitons ($A^0$ and $D^0$) whose recombination requires no defect mediation [Figs. 2(a) and 2(e)].

Beside the defect-mediated recombination, the intervalley excitons can also decay through emission of zone-corner chiral phonons.    We have observed a peak ($I_p^0$) at ~26.6 meV below $I^0$ (Fig. 2).    We assign $I_p^0$ as a chiral-phonon replica of $I^0$ based on four observations.    First, $I_p^0$ shows parallel gating dependence to $I^0$ in PL energy, intensity and linewidth [Figs. 2(c)-2(e)].    Second, the $I_p^0$ $g$ factor (12.5) closely matches the $I^0$ $g$ factor (12.4) [Figs. 3(c) and 3(d)].    Third, the redshift energy (26.6 meV) and intervalley momentum difference match the energy and momentum of the zone-corner chiral phonons in monolayer WSe$_2$ [66, 67].    Finally, the $I_p^0$ optical helicity matches the selection rules predicted by prior studies [40, 68].

The $I_p^0$ selection rules can be illustrated by consideirng the decay of an intervalley electron-hole pair into a photon and a zone-corner phonon [Fig. 1(b)].    The transition rate according to the second-order perturbation theory is [28, 60]:

$$P = \frac{2\pi}{\hbar} \left| \frac{\langle v_{\uparrow K} | \hat{H}_{el} | c_{2,\uparrow K} \rangle \langle c_{2,\uparrow K} | \hat{H}_{ep} | c_{1,\uparrow K'} \rangle}{E_{c1} - E_{c2} - \hbar\Omega} \right|^2 \delta(E_{c1} - E_v - \hbar\Omega - \hbar\omega). \tag{1}$$

Here $\hat{H}_{el}$ ($\hat{H}_{ep}$) is the electron-light (electron-phonon) interaction Hamiltonian; $\hbar\omega$ ($\hbar\Omega$) is the photon (phonon) energy.    $c_{2,\uparrow K}$ at the K point is the dominant mediating state.    As the $\hat{H}_{el}$ matrix element represents intravalley transition, the intervalley $I_p^0$ emission will follow the same optical selection rules as the intravalley $A^0$ exciton – the emitted photon will be right-handed ( $\sigma^+$ ) [28, 40, 60, 68, 69].    According to the irreducible representations of the electronic states at the K and K' points [shown in Fig. 1(b)], the emitted K'-point phonon will have the $K_3$ representation by group theory – it is a left-handed chiral phonon ($\Omega^-$) [40, 67, 68] (see Sec. 8 in Supplemental Material [27]).    The time-reversal intervalley exciton will have opposite photon and phonon chirality.    These selection rules are fully consistent with our experimental results (Fig. 3).

The selection rules of the defect-mediated $I^0$ emission can be similarly derived from Eq. (1) by replacing $\hat{H}_{ep}$ with a defect interaction Hamiltonian and setting $\hbar\Omega = 0$.    The defect potential can couple $c_{1,\uparrow K'}$ to $c_{2,\uparrow K}$ by breaking the translational and rotational symmetry of the crystal.    As light emission is governed by the same $\hat{H}_{el}$ matrix element, $I^0$ will have the same optical helicity as $I_p^0$ and $A^0$, consistent with our observations [Figs. 3(c) and 3(d)].

With a good understanding of $I^0$ and $I_p^0$ at charge neutrality, we turn to the $D^+$, $D^-$ and $I_p^+$, $I_p^-$ peaks on the hole and electron sides.    Previous study has proved that $D^+$ and $D^-$ are the dark trion emission [29, 70].    Here we identify $I_p^+$ and $I_p^-$ as the zone-corner chiral-phonon replicas of dark trion emission based on three observations.    First, $I_p^+$ ($I_p^-$) shows similar gate dependence to $D^+$ ($D^-$) in PL energy, intensity, and linewidth (Fig. 2).     Second, $I_p^+$ $(I_p^-)$ is 26.5±0.5 meV below $D^+$ $(D^-)$, consistent with the $I_p^0 - I^0$ separation (26.6 meV).     Finally, $I_p^+$, $I_p^-$ exhibit the same optical helicity as the intravalley bright states ($A^0$, $A^+$, $A_1^-$, and $A_2^-$) (Fig. 3); this is consistent with the selection rules of chiral-phonon replicas by group theory.

Unlike $I_p^0$ and $I^0$ that have practically the same g factor, however, $I_p^+$ and $I_p^-$ have significantly larger g factors ($g = 13.7, 11.9$, respectively) than $D^+$ and $D^-$ ($g = 10.4, 9.9$, respectively) (Fig. 3).     This difference indicates that $I_p^+$ and $I_p^-$ are not truly the replicas of $D^+$ and $D^-$.     To resolve this conflict, we need to consider that a trion can decay in two paths – intravalley or intervalley electron-hole recombination.     The intravalley transitions produce the previously studied $D^+$ and $D^-$ lines [29, 70].     The intervalley transitions (labeled as $I^+$ and $I^-$) are hitherto unreported.     Our observed $I_p^+$ and $I_p^-$ lines should be the replicas of $I^+$ and $I^-$ (not $D^+$ and $D^-$), because zone-corner phonons are associated with intervalley transitions.

We note that $I^+$ and $I^-$ emission may be mediated by defect scattering.     At zero magnetic field, $I^+$ $(I^-)$ is energy-degenerate with $D^+$ $(D^-)$.     After a trion decays through intra- or inter-valley recombination, the final state is a free carrier in either valley. As the valleys are degenerate, both transitions should have the same energy.     This contrasts with the different energy between the intra- and inter-valley excitons ($D^0$ and $I^0$).

We can distinguish the intra- and inter-valley trion emission by applying a strong magnetic field to lift the valley degeneracy.     Under positive magnetic field, the K' valley has a larger band gap than the K valley [30-34].     The initial fourfold degenerate trion emission is split into four distinct lines, labeled as $I_K^+$, $D_{K'}^+$, $D_K^+$, and $I_{K'}^+$ [Figs. 4(a) and 4(b)].     Figure 4(c) displays the left-handed PL map for positive trions at $B = 15$ T (See Fig. S10 for the results of right-handed helicity and negative trions [27]).     We observe the $I_K^+$, $D_{K'}^+$, and $D_K^+$ lines; The $I_{K'}^+$ line with right-handed helicity is absent.     The intra- and inter-valley peaks ($D_{K'}^+$ and $I_K^+$) continuously separate from each other under an increasing magnetic field [Fig. 4(d)].     Their energy difference can be fit linearly by $\Delta E = g\mu_B B$ with $g = 2.35 \pm 0.1$ [Fig. 4(e)].     Notably, here $\Delta E$ is identical to the energy separation between the K and K' points of the lower conduction band ($c_1$), which is $\Delta E = (2m_0/m_e^* - 2)\mu_B B$ in a single-particle model ($m_0$ and $m_e^*$ are the free and effective electron mass, respectively).     From our measured g factor, we deduce $m_e^* = 0.46 \pm 0.01m_0$ [27], consistent with the results from first-principle calculations [37].     The result thus confirms the intervalley trion transition. This is the first measurement of the effective mass of the lower conduction band in monolayer WSe$_2$.     This effective mass, obtained from trions, may differ slightly from the effective mass at charge neutrality due to the trion-lattice coupling [52].

After we separate $I_K^+$ from $D_{K'}^+$ under a magnetic field, it becomes clear that $I_p^+$ is the zone-corner phonon replica of $I_K^+$, because their energy separation (~26.5 meV) remains unchanged under a magnetic field [Fig. 4(c)].   $D_p^+$ is also proven to be the zone-center chiral-phonon replica of $D_{K'}^+$ with ~21.5 meV separation [28, 60].   Therefore, the intra- and inter-valley trion transitions are coupled selectively to the zone-center and zone-corner chiral phonons, respectively [Figs. 1(c) and 1(d)].

Our results show that the same initial trion state in monolayer $WSe_2$ can decay through four types of quantum paths, including the (i) intravalley transition; (ii) intervalley transition; (iii) intravalley transition with zone-center chiral-phonon emission; and (iv) intervalley transition with zone-corner chiral-phonon emission.   Such multipath emission is much more intricate than the usual single-path trion emission in prior research.   The interplay of intra- and inter-valley trion transitions can strongly enrich the quantum dynamics in trions.   More generally, by leveraging the multipath transitions in intervalley excitons and trions, we may enable sophisticated control of excitonic dynamics, probe the condensation of intervalley excitons, and explore novel chiral-phonon phenomena.

**Acknowledgement:** We thank Z. Lu and D. Smirnov for assistance in magneto-optical experiments and H. Dery for fruitful and insightful discussions on the interpretation of the data.   Y.C.C. is supported by Ministry of Science and Technology (Taiwan) under Grants No. MOST 107-2112-M-001-032 and No. 108-2112-M-001-041.   N.M.G. acknowledges support from the National Science Foundation Division of Materials Research CAREER Grant No. 1651247.   K.W. and T.T. acknowledge support from the Elemental Strategy Initiative conducted by the MEXT, Japan and the CREST (JPMJCR15F3), JST.   A large portion of this work was performed at the National High Magnetic Field Laboratory, which is supported by the National Science Foundation Cooperative Agreement No. DMR-1644779 and the State of Florida.

E. L and J. v. B. contributed equally to this work.

*Note added.* --- After we submitted our manuscript, two related papers were published [72, 73].

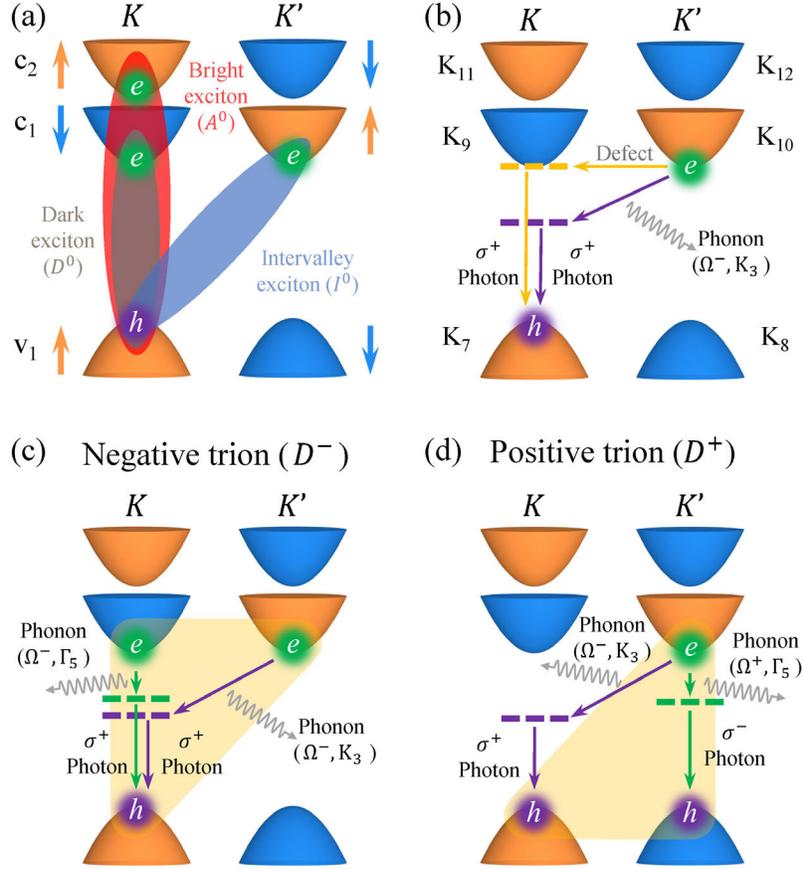

FIG. 1. Intervalley electron-hole recombination processes in monolayer WSe$_2$. (a) Band configurations for the intravalley bright exciton ($A^0$), intravalley dark exciton ($D^0$), and intervalley momentum-forbidden dark exciton ($I^0$). The arrows and color denote the electron spin. (b) Recombination of $I^0$ mediated by defects or zone-corner chiral phonons. We denote the irreducible representations of the electronic states at the K and K' points with the $C_{3h}$ point group. (c), (d) Trion decay through emission of zone-center or zone-corner chiral phonons. $\Gamma_5$ (K$_3$) is the representation of chiral phonons at the $\Gamma$ (K') point in the $D_{3h}$ ($C_{3h}$) point group. $\Omega^{\pm}$ and $\sigma^{\pm}$ denote the chirality of the emitted phonons and photons, respectively.

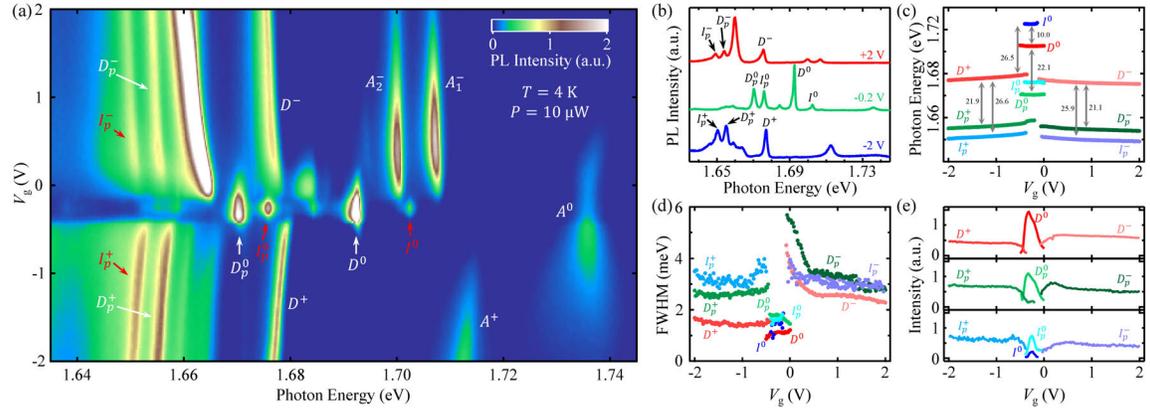

FIG. 2. Gate-dependent PL of monolayer $WSe_2$. (a) PL map at $T \sim 4$ K and zero magnetic field under 532-nm continuous laser excitation with incident laser power $P \sim 10 \, \mu W$. We denote the bright exciton and trions ($A^0$, $A^+$, $A_1^-$, and $A_2^-$), dark exciton and trions ($D^0$, $D^+$, and $D^-$), intervalley dark exciton ($I^0$), zone-center chiral-phonon replicas of dark states with $\sim 21.4$ meV redshift ($D_p^0$, $D_p^+$, and $D_p^-$), and zone-corner chiral-phonon replicas of dark states with $\sim 26.5$ meV redshift ($I_p^0$, $I_p^+$, and $I_p^-$). (b) Cross-cut PL spectra on the electron side, charge-neutrality point, and the hole side with respective gate voltages $V_g = +2$, $-0.2$, $-2$ V. (c)-(e) The PL energy, full width at half maximum (FWHM), and integrated PL intensity of different excitonic peaks as a function of gate voltage. The numbers in (c) denote the energy separation in unit of meV.

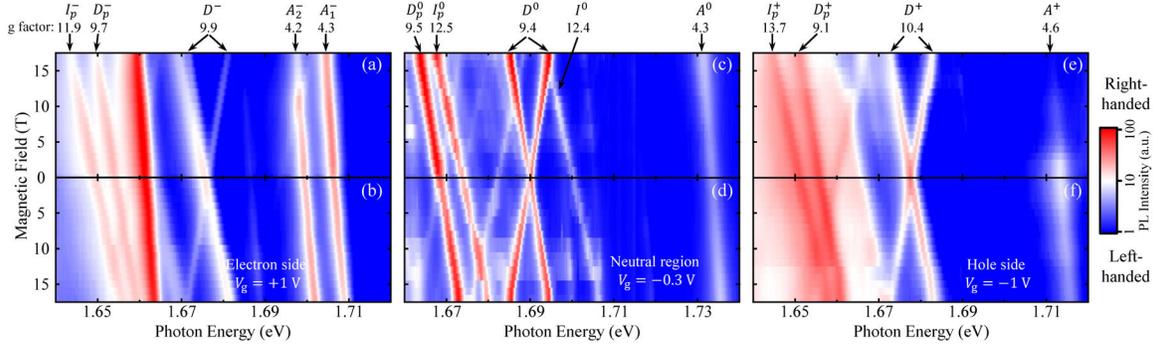

FIG. 3. Magnetic-field-dependent and helicity-resolved PL in monolayer WSe₂. (a), (b) PL maps at magnetic fields $B = 0 - 17$ T on the electron side. (c), (d) Similar PL maps near the charge neutrality point. (e), (f) Similar PL maps on the hole side. The top and bottom rows show PL with right- and left-handed helicity, respectively. The emission lines and their $g$ factors are denoted on the top.

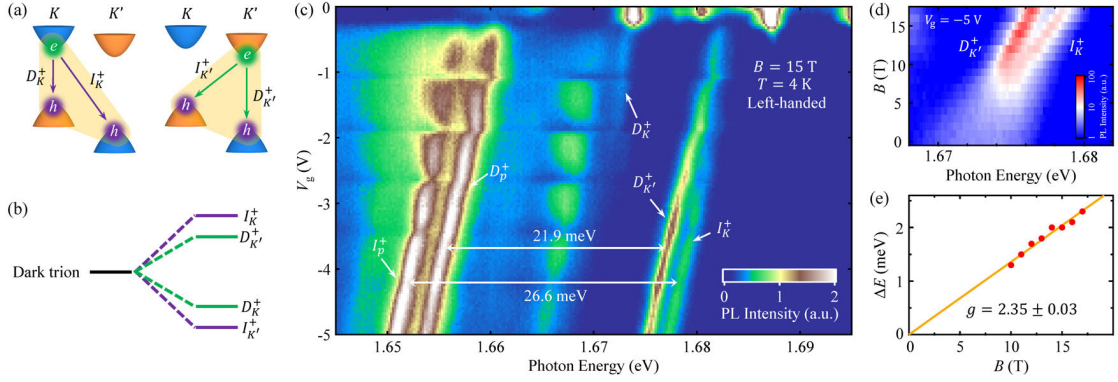

FIG. 4. Splitting of intra- and inter-valley trion emission under magnetic field. (a), (b) Schematic configurations and energy levels of different trion transitions. The application of a magnetic field splits the fourfold degenerate trion emission into four distinct emission lines, labeled as $I_K^+$, $D_{K'}^+$, $D_K^+$, and $I_{K'}^+$ from high to low energy. Here $D$ and $I$ denote intra- and inter-valley transitions, respectively; the K and K' subscripts denote the valley of the electron in trion. (c) The left-handed PL map on the hole side of monolayer WSe$_2$ at $B$ = 15 T. The different transitions and energy separations are denoted. The gate-dependent oscillations of PL intensity come from Landau quantization [39, 71]. (d) Log-scale magnetic-field-dependent PL map of positive trions at $B$ = 0 – 17 T. (e) Energy separation between the two split trion peaks in (d) as a function of the magnetic field. The line is a linear fit that gives an effective $g$ factor 2.35 ± 0.03.

detailed description of experimental conditions, additional data, calculations and Refs. [19, 28-52].

Supplemental Material of

**"Multipath optical recombination of intervalley dark excitons and trions in monolayer WSe₂"**

Erfu Liu[1], Jeremiah van Baren[1], Ching-Tarng Liang[2], Takashi Taniguchi[3], Kenji Watanabe[4], Nathaniel M. Gabor[1,5], Yia-Chung Chang[2], Chun Hung Lui[1*]

[1] Department of Physics and Astronomy, University of California, Riverside, CA 92521, USA.

[2] Research Center for Applied Sciences, Academia Sinica, Taipei 11529, Taiwan.

[3] International Center for Materials Nanoarchitectonics (WPI-MANA), National Institute for Materials Science (NIMS), 1-1 Namiki Tsukuba, Ibaraki 305-0044, Japan.

[4] National Institute for Materials Science (NIMS), 1-1 Namiki Tsukuba, Ibaraki 305-0044, Japan.

[5] Canadian Institute for Advanced Research, MaRS Centre West Tower, 661 University Avenue, Toronto, Ontario ON M5G 1M1, Canada.

[*]Corresponding author. Email: joshua.lui@ucr.edu

# Table of contents

# 1. Device fabrication

We fabricate monolayer WSe$_2$ gating devices encapsulated by hexagonal boron nitride (BN) by mechanical co-lamination of two-dimensional (2D) crystals. We use WSe$_2$ bulk crystals from HQ Graphene Inc. We first exfoliate monolayer WSe$_2$, multi-layer graphene, and thin BN flakes from their bulk crystals onto Si/SiO$_2$ substrates. Afterward, we apply a polycarbonate-based dry-transfer technique to stack these different 2D crystals together. We use a stamp to first pick up a BN flake, and sequentially pick up multi-layer graphene (as the electrodes), a WSe$_2$ monolayer, a BN thin layer (as the bottom gate dielectric), and multi-layer graphene (as the back-gate electrode). This method ensures that the WSe$_2$ layer doesn't contact the polymer during the whole fabrication process to reduce the contaminants and bubbles at the interface. Afterward, we apply the standard electron beam lithography to pattern and deposit the gold contacts (100 nm thickness). Finally, we anneal the devices at 300 ℃ for 5 hours in an argon environment to cleanse the interfaces.

Figure. S1(a-b) display the schematic of our devices and the optical image of a representative device. In this paper, all the results were obtained on Device 1, except some results in Fig. S5 that were obtained on Devices 2, 3, 4. For Device 1, the BN dielectric layer of the back gate is ~42 nm in thickness, from which we can deduce the capacity and injected charge density of the back gate.

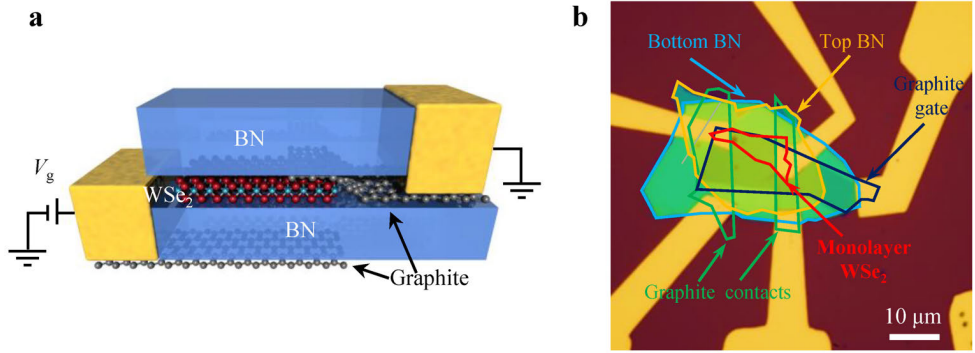

**Fig. S1. (**a) The schematic of BN-encapsulated monolayer WSe$_2$ device. (b) The optical image of a representative device.

# 2. Experimental methods

We measured Device 1 in the National High Magnetic Field Laboratory (NHMFL) in Tallahassee, Florida, United States. We made two trips to NHMFL to finish our measurements. All the data taken on Device 1, except Fig. 3(c-d) in the main paper, was obtained during the first trip in August 2019. The data in Fig. 3(c-d) was obtained on Device 1 during the second trip in January 2020. There is a small energy shift (~2 meV) between the spectra obtained in the first and second trips due to systematic error, but this small energy shift does not affect our conclusion.

Our experiments in NHMFL were conducted with a 17.5-T DC superconducting magnet (Cell 3). Fig. S2 displays the schematic of the experimental setup. The sample is mounted on a three-dimensional piezoelectric translation stage. The sample temperature is T ~ 4 K in our experiment. A linearly polarized 532-nm continuous laser is focused onto the sample with a spot diameter of ~2 μm by an objective lens (NA = 0.6). The photoluminescence (PL) is collected by the same objective, passes through a 50/50 beam splitter into a multimode optical fiber, and is subsequently measured by a spectrometer with a CCD camera (Princeton Instruments, IsoPlane 320). A quarter wave plate and a polarization beam splitter are inserted in the collection beam path to select the right-handed or left-handed circularly polarized component of the PL signals.

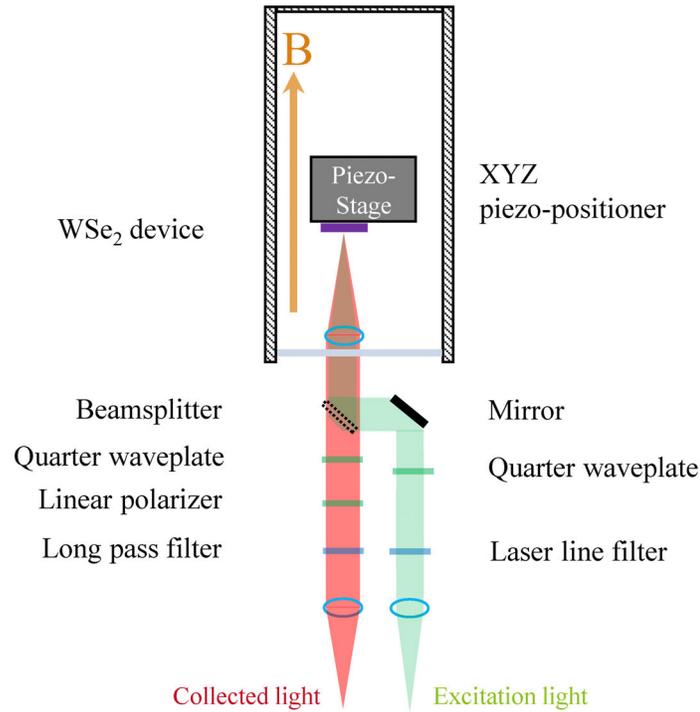

**Fig. S2.** Schematic of our magneto-optical experimental setup.

Our measurement geometry has different PL collection efficiency for the bright and dark excitonic states, which are associated with in-plane and out-of-plane dipole, respectively. By assuming a $\sin^2 \theta$ radiation intensity pattern of an oscillating dipole, our microscope objective (NA = 0.6) can collect ~7% radiation from the dark excitonic states with out-of-plane dipole and ~60% radiation from bright excitonic states with in-plane dipole, including the reflected light from the substrate. So, the collection efficiency for dark states is about ten times smaller than that for bright states. Moreover, the oscillator strength of dark excitons is ~400 times smaller than that of bright excitons, according to our previous calculation [1]. However, the dark-state lifetime (>100 ps) is much longer than the bright-state lifetime (<10 ps) [2]. The dark-state population is also much larger than the bright-state population at low temperature because the dark states lie at the lowest energy. The longer lifetime and larger population can strongly enhance the dark-state PL. These factors can account for the comparable PL intensity of the bright and dark states observed in our experiment.

Our measurement geometry favors the collection of PL from the phonon replicas. The phonon replicas are expected to have smaller oscillator strength than the original dark states [1]. But the collection efficiency of phonon replicas is ~60% because they are associated with in-plane dipole (like the bright exciton), much higher than the ~7% collection efficiency of the dark exciton. So, the phonon replicas appear to be strong in our experiment.

## 3. Characterization of monolayer WSe₂ devices

### 3.1. Photodoping effect

The performance of monolayer WSe₂ devices can be affected by the photodoping effect. The photodoping effect is typically manifested by the hysteresis when the gate voltage is swept back and forth. To this end, we have measured the PL maps of Device 1 by sweeping the gate voltage from –2 V to +2 V and from +2 V to –2V (Fig. S3). We do not observe any noticeable hysteresis between them. The result indicates negligible photodoping effect in the device.

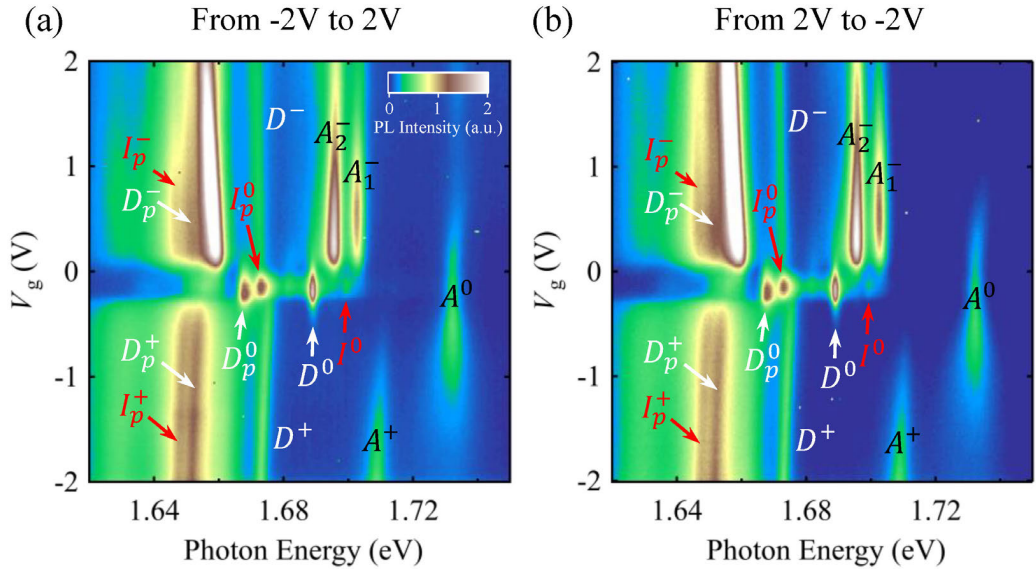

**Fig. S3.** (a-b) PL maps of monolayer WSe₂ (Device 1) with the gate voltage ($V_g$) swept from –2 V to +2 V (a) and from +2 V to –2 V (b). We do not observe any noticeable hysteresis between them, indicating negligible photodoping effect in the device. The PL is measured under 532-nm continuous laser excitation at temperature T ~ 5 K. The incident laser power is ~10 μW.

### 3.2. Influence of excitation photon energy

We have measured the PL maps of monolayer WS₂ (Device 1) under the excitation of two different lasers – a 532-nm green laser (photon energy 2.33 eV) and a 633-nm red laser (photon energy 1.96 eV) (Fig. S4). The $I^0$ intervalley exciton emission line and its phonon replica $I_p^0$ are observed in both excitation photon energies. Therefore, the intervalley emission is robust against a modest change of excitation photon energy.

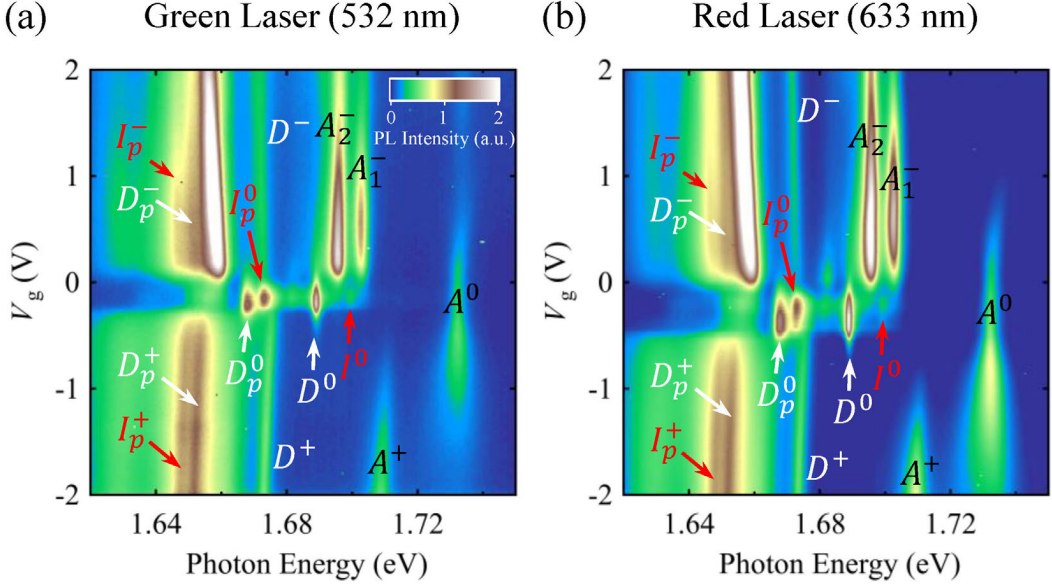

**Fig. S4.** (a-b) Gate-dependent PL maps of monolayer WSe₂ (Device 1) under the excitation of 532-nm green laser (a) and 633-nm red laser (b).

### 3.3. Reproducibility of the results in different devices

Besides Device 1, whose results are presented in the main paper, we have also measured three other BN-encapsulated monolayer WSe₂ devices (Devices 2, 3, 4). The experiments on Devices 2 – 4 were conducted with an optical cryostat with no magnetic field (Montana Instruments) in our laboratory at UC Riverside. The sample temperature is roughly estimated to be ~15 K.

Fig. S5(a-d) compare the gate-dependent PL maps of Devices 1 – 4 under 532-nm continuous laser excitation at low temperature with no magnetic field. All of the four devices exhibit the intervalley excitonic emission and the associated phonon replicas. Their results are consistent with each other. The main paper presents the results of Device 1, which shows sharper emission lines than the other devices.

For a finer comparison, we plot the cross-cut spectra near the charge neutrality point of these four maps in Fig. S5(e). The absolute photon energies of the emission peaks are somewhat different in these four devices. Such variation of emission photon energies in different devices may come from their different BN thickness and different TMD strain induced in the fabrication process. For instance, the BN thickness can affect the dielectric screening; the TMD strain can modulate the electronic structure. Both can modulate the exciton emission energy. The absolute photon energies of the exciton peaks can therefore vary from device to device.

However, the energy separation between different peaks are largely consistent in these four devices. In Fig. S5 (f), we shift the spectra horizontally to align with respect to the $I^0$ position. We see that the energy separation between $I^0$ and $D^0$, $I_p^0$, $D_p^0$ are almost the same in these four devices. Overall, the essential features of the intervalley exciton and replica are consistent and reproducible in Devices 1 – 4.

We find that the $I^0 - A^0$ energy separation is somewhat larger in Device 1 than in Devices $2 - 4$, in contrast to the $I^0 - D^0$ separation that is almost the same in all four devices. The results suggest that the binding energies of $I^0$ and $A^0$ change differently with the sample conditions (*e.g.* TMD strain and BN thickness), whereas the binding energies of $I^0$ and $D^0$ change similarly with the sample conditions. We can explain such difference from their different characteristics. $I^0$ and $A^0$ have different binding energies because the upper and lower conduction bands have different electron effective mass. The absolute change of the $I^0$ and $A^0$ binding energies should be different upon a change of sample conditions, if we assume similar fractional change of their binding energies. In contrast, $I^0$ and $D^0$ have the same effective mass. Their binding energies differ only by the electron-hole exchange energy ($\sim$10 meV). As the electron-hole exchange energy is not sensitive to the sample conditions, the absolute change of the $I^0$ and $D^0$ binding energies should be similar upon a change of sample conditions. This can account for the different sample dependence of the $I^0 - A^0$ separation and $I^0 - D^0$ separation, as observed in our experiment.

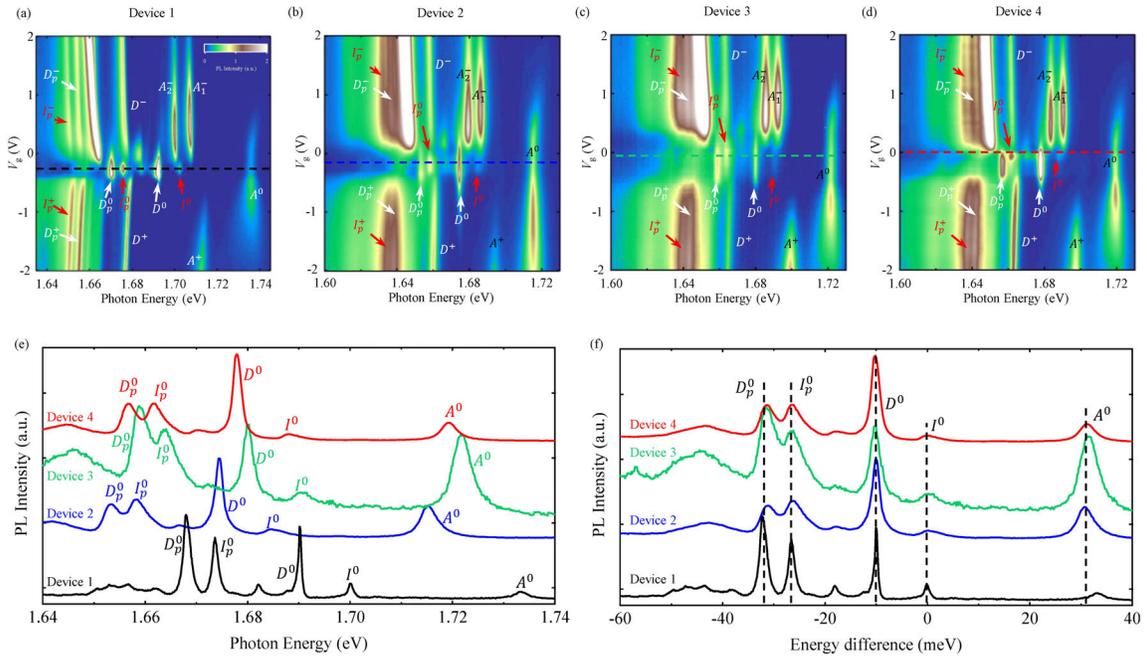

**Fig. S5.** (a-d) Comparison of gate-dependent PL maps in four BN-encapsulated monolayer WSe$_2$ devices (Devices $1 - 4$) under 532-nm continuous laser excitation with no magnetic field. The incident laser power is $\sim$10 μW in all the measurements. The sample temperature is $\sim$ 4 K for Device 1 and $\sim$ 15 K for Devices 2, 3, 4. The four maps (a-d) share the same color scale bar in (a), but have different arbitrary units. The different emission lines are denoted. (e) Cross-cut PL spectra near the charge neutrality in (a-d) (marked by the dashed lines). (f) The same spectra in (e) as a function of energy separation from the $I^0$ peak.

## 4. Power dependence of photoluminescence

We have measured the PL from the intravalley bright excitons ($A^0$) and dark excitons ($D^0$), intervalley excitons ($I^0$), and the phonon replicas ($D_p^0$, $I_p^0$) at different laser power ($P$) for Device 1 [Fig. S6(a)]. Their PL intensity ($I$) all exhibits approximately linear power ($P$) dependence. The power-law fits $I = P^\gamma$ give the exponents $\gamma = 1.04, 0.87, 0.95, 0.88$ and $0.85$ for the $A^0$, $D^0$, $I^0$, $D_p^0$ and $I_p^0$ peaks, respectively. The nearly linear power dependence shows that they are not associated with the biexciton states with quadratic power dependence. Fig. S6(b) shows their full width at half maximum (FWHM) as a function of incident laser power. The FWHM increases with the laser power. The data in the main paper was taken at low incident laser power (~10 µW) to sharpen the emission lines.

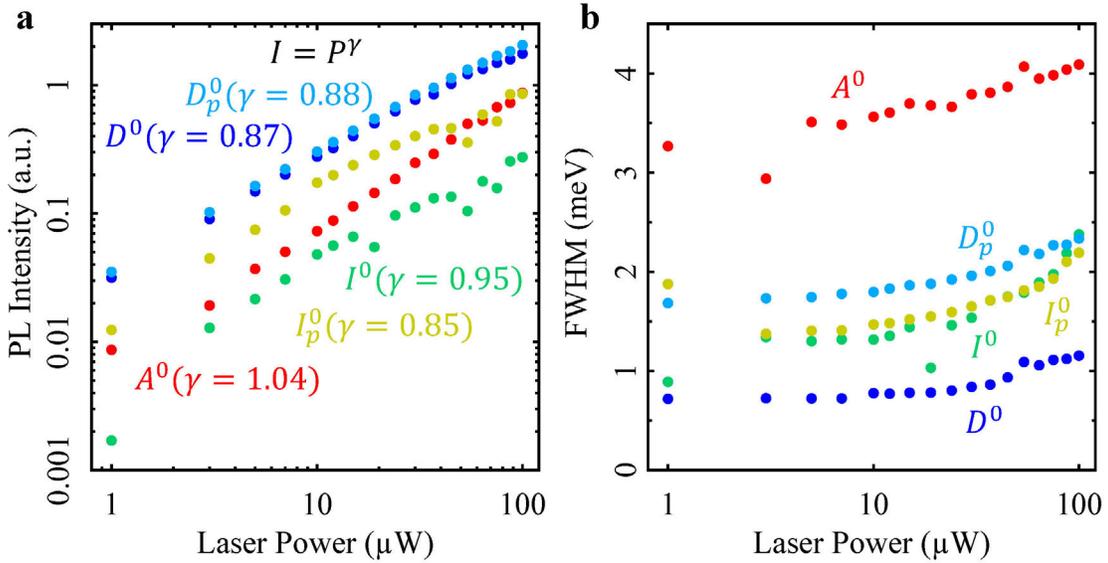

**Fig. S6.** (a) Photoluminescence (PL) intensity as a function of incident laser power for the bright exciton ($A^0$), dark exciton ($D^0$) and its phonon replica ($D_p^0$), intervalley exciton ($I^0$) and its phonon replica ($I_p^0$). The lines are fits by the power-law $I = P^\gamma$. (b) Full width at half maximum (FWHM) of these peaks as a function of incident laser power.

## 5. Energy seperation between phonon replicas and their original lines

We have extracted the peak energies of dark exciton and trions ($D^0$, $D^-$, $D^+$), intervalley dark exciton ($I^0$), and the related phonon replicas ($D_p^0$, $D_p^-$, $D_p^+$, $I_p^0$, $I_p^-$, $I_p^+$) at zero magnetic field for Device 1. Fig. S7 displays the energy separation between the phonon replicas and their primary emission lines. The replica redshift energies remain almost unchanged in the whole gate range. We notice slightly smaller replica energies on the electron side (positive gate voltage) than in the charge-neutral regime and on the hole side (negative gate voltage). The deviation implies somewhat asymmetric photon-carrier interactions on the electron and hole sides. Further research is merited to clarify this issue.

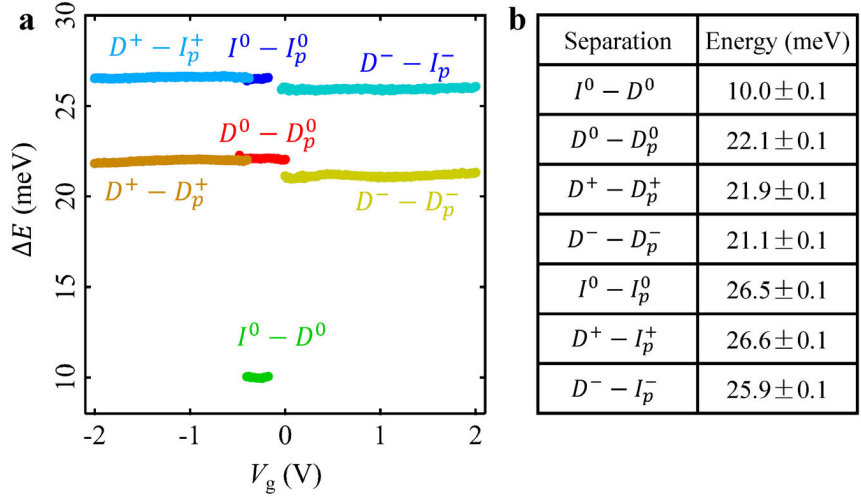

**Fig. S7.** (a) Energy seperation between the phonon replicas and their original emission lines. (b) List of the average energy seperation as shown in panel a.

## 6. Zeeman effect and g-factors

The degeneracy of the K and K' valleys in monolayer WSe₂ can be lifted by breaking the time-reversal symmetry with out-of-plane magnetic field (B). The magnetic field can increase the energy gap of one valley and diminish the energy gap of the other valley. The difference between the two valley gaps is defined as the valley Zeeman splitting energy $\Delta E = g\mu_B B$, where $\mu_B$ is the Bohr magneton and $g$ is the effective g-factor.

In a single-particle model, the Zeeman shift of a band in monolayer WSe₂ is [3-8]:

$$E_Z = \left(2s + m\tau + \frac{m_0}{m^*}\tau\right)\mu_B B. \tag{S1}$$

The first term is the spin Zeeman shift. $s = \pm 1/2$ is the spin quantum number of the band. The second term is the atomic-orbit Zeeman shift. $\tau$ is the valley index for the K ($\tau = +1$) and K' ($\tau = -1$) valleys. $m$ is the atomic-orbit azimuthal quantum number in the conduction band ($m = 0$) and valence band ($m = 2$). The third term is the Berry-curvature contribution. $m_0$ is the free electron mass and $m^*$ is the effective carrier mass. In monolayer WSe₂, the upper conduction band (c₂) and the valence band (v₁) have similar carrier effective mass $m^* \sim 0.42m_0$ [9-11]. The effective mass of the lower conduction band is $m^* \sim 0.46m_0$ accroding to our trion results in the main text.

For the intravalley excitons ($A^0$, $D^0$), the Berry-curvature contributions from the conduction and valence bands almost completely cancel each other. The $A^0$ and $D^0$ excitons with parallel and antiparallel electron spins have predicted g-factors $g_A \approx 4$ and $g_D \approx 8.4$, respectively (Fig. S8). In contrast, for the intervalley exciton ($I^0$), the Berry-curvature contributions from the conduction and valence bands add up. The intervalley exciton has a predicted g-factor $g_I \approx 13.1$ (Fig. S8).

In experiment, the Zeeman splitting energy can be measured either by the exciton PL from opposite valleys under the same magnetic field or by the exciton PL of one valley under opposite magnetic fields [3-6]. Fig. 3 of the main text shows our measured helicity-resolved PL maps under continuous magnetic field. Here in Fig. S9, we further show the fitted energies of different emission lines. The measured Zeeman-splitting g-factor is ~4 for the intravalley bright states, ~9 for the intravalley dark states and replicas, and ~13 for the intervalley states and replicas. These experimental g-factors agree with the predicted values from the single-particle picture.

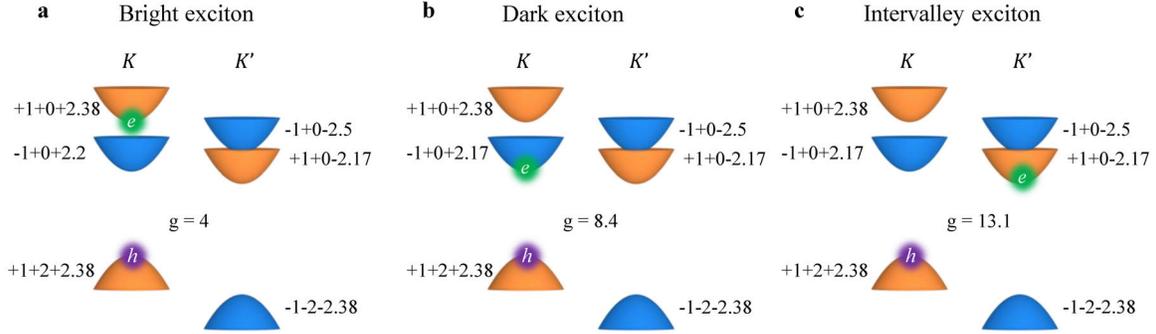

**Fig. S8.** The Zeeman splitting g-factors for the intravalley bright exciton, intravalley dark exciton, and intervalley exciton, predicted by the single-particle model described in the text. The Zeeman-shift g-factor of a band have three contributions from the spin, atomic orbit, and Berry curvature, whose component g-factor is denoted, respectively, by the first, second and third numbers in the figures. The sum of these three numbers is the g-factor of the band. The difference between the g-factor of a conduction band and a valence band is the Zeeman-shift g-factor of the exciton. The difference of the exciton g-factors between the two valleys is the Zeeman-splitting g-factor of the exciton (which is denoted in the middle of each panel).

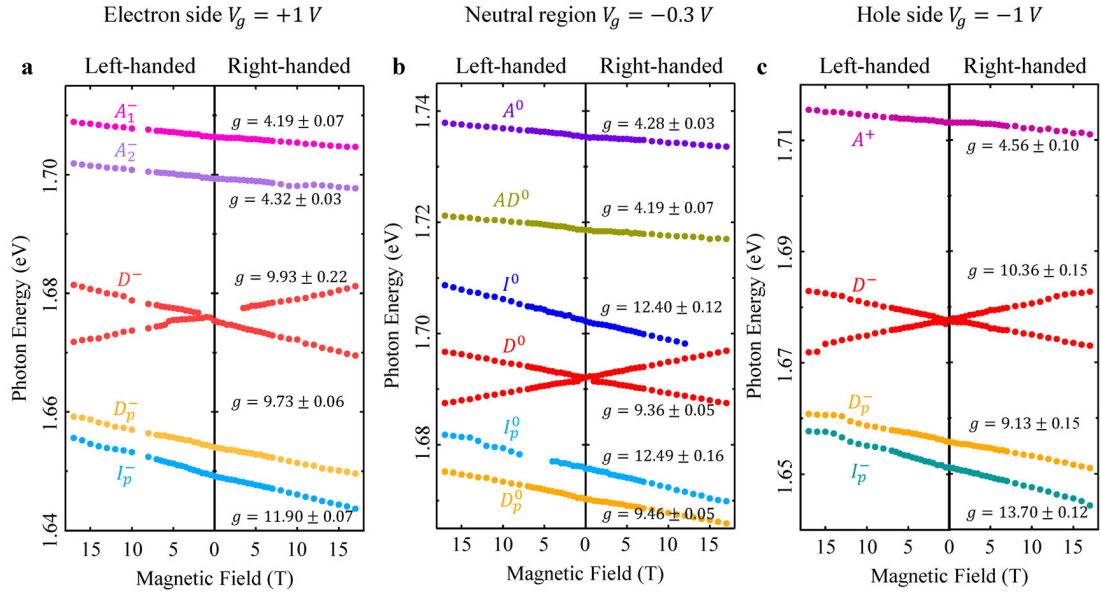

**Fig. S9.** (a-c) Magnetic-field dependent PL peak energies of monolayer WSe$_2$ on the electron side (a; $V_g = 1$ V), near the charge neutrality point (b; $V_g = -0.3$ V) and on the hole side (c; $V_g = -1$ V). We excite the sample with linearly polarized 532-nm laser and collect the PL with right-handed helicity (right halves) and left-handed helicity (left halves). The g-factors of different peaks are denoted.

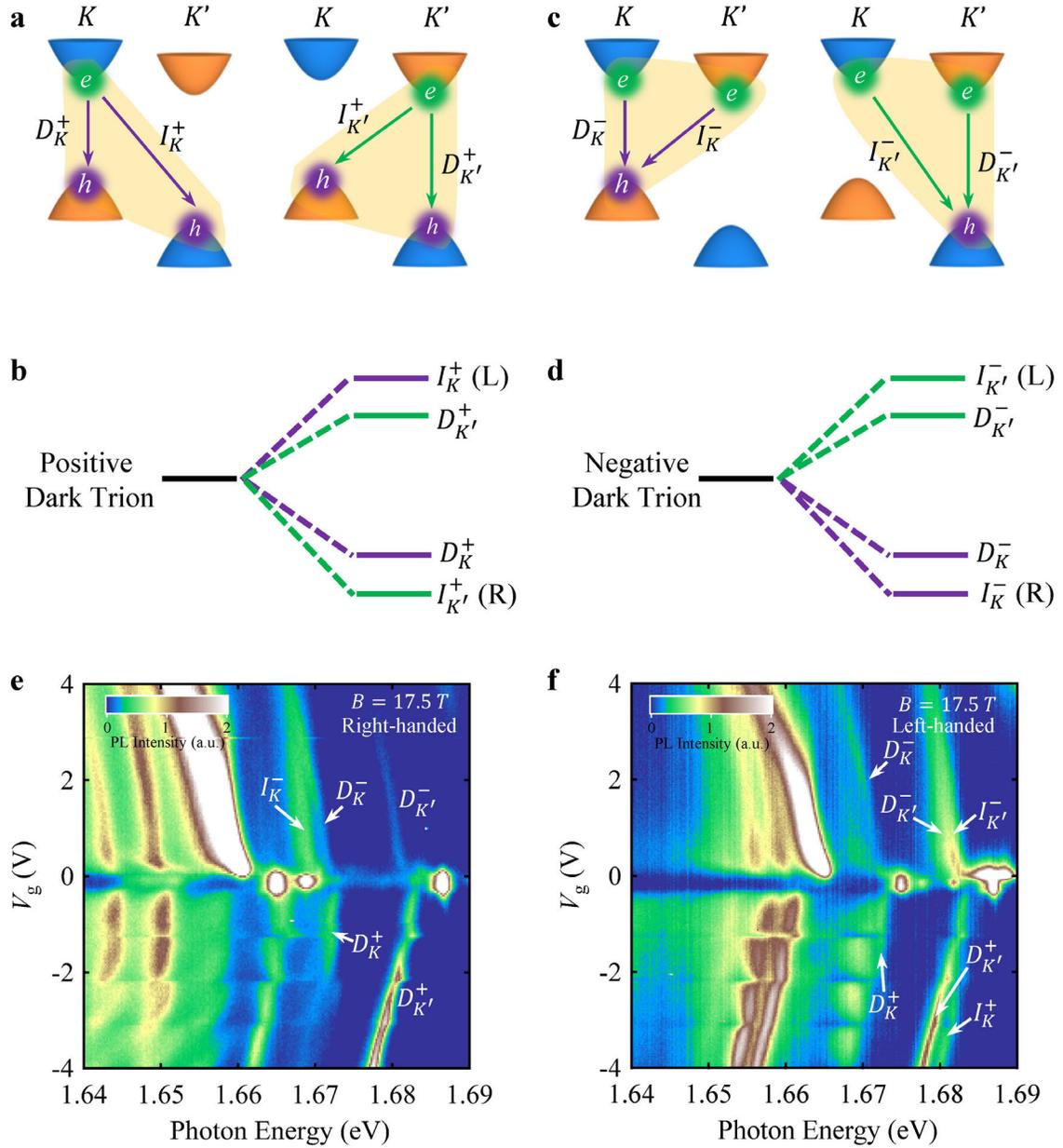

**Fig. S10** (a-b) Different recombination processes and relative energy levels for the positive dark trions in monolayer WSe₂ under magnetic field. The four transition processes are labeled as $I_K^+$, $D_{K'}^+$, $D_K^+$, $I_{K'}^+$, where the K and K' subscripts denote the valley of the electron and "D" and "I" denote intra- and inter-valley transitions, respectively. (c-d) Similar diagrams for negative dark trions, with the K and K' subscripts denoting the valley of the hole. "L" and "R" denote left-handed and right-handed optical helicity determined by group theory. (e-f) Gate-dependent PL maps of monolayer WSe₂ at right-handed and left-handed helicity at B = 17.5 T. The trion emission peaks are denoted.

## 7. Splitting of dark trion emission under magnetic field

Figure 4 in the main paper shows the left-handed PL map on the hole side for monolayer WSe$_2$ at B = 15 T. Here in Fig. S10, we show the complete PL maps in both left-handed and right-handed helicity and for both the positive and negative trions. Fig. S10(a-d) illustrate the four different recombination processes and their relative energy levels for positive and negative trions under magnetic field. The intravalley emission has linear polarization. The intervalley emission has circular polarization; the helicity determined by group theory is denoted in Fig. S10(b, d).

For the negative trions, we can resolve three emission lines ($I_K^-$, $D_K^-$, $D_{K'}^-$) in the right-handed helicity [Fig. S10(e)]. The $I_{K'}^-$ line is absent due to its left-handed helicity. On the other hand, we also resolve three emission lines ($I_{K'}^-$, $D_K^-$, $D_{K'}^-$) in the left-handed helicity [Fig. S10(f)]. The $I_K^-$ line is absent due to its right-handed helicity. In sum, we observe all four trion emission lines on the electron side, and also confirm the helicity of the two intervalley emission lines.

For the positive trions, we can resolve three emission lines ($D_K^+$, $D_{K'}^+$, $I_K^+$) in the left-handed helicity (presented in the main text), but only two emission lines ($D_K^-$, $D_{K'}^-$) in the right-handed helicity due to the broad and weak optical features [Fig. S10(e)-S10(f)]. Further improvements of devices and experiment are needed to resolve the missing trion line. We note that pronounced quantum oscillations are observed in the PL maps. These oscillations are induced by the formation of Landau levels under magnetic field [12].

## 8. Irreducible representations of zone-center and zone-corner states

| $D_{3h}$ | $E$ | $\bar{E}$ | $C_3^+ C_3^-$ | $\bar{C}_3^+ \bar{C}_3^-$ | $\sigma_h \bar{\sigma}_h$ | $S_3^+ S_3^-$ | $\bar{S}_3^+ \bar{S}_3^-$ | $C'_{2i} \bar{C}'_{2i}$ | $\sigma_{vi} \bar{\sigma}_{vi}$ |
|---|---|---|---|---|---|---|---|---|---|
| $A_1' \, \Gamma_1$ | 1 | 1 | 1 | 1 | 1 | 1 | 1 | 1 | 1 |
| $A_2' \, \Gamma_2$ | 1 | 1 | 1 | 1 | 1 | 1 | 1 | -1 | -1 |
| $A_1'' \, \Gamma_3$ | 1 | 1 | 1 | 1 | -1 | -1 | -1 | 1 | -1 |
| $A_2'' \, \Gamma_4$ | 1 | 1 | 1 | 1 | -1 | -1 | -1 | -1 | 1 |
| $E'' \, \Gamma_5$ | 2 | 2 | -1 | -1 | -2 | 1 | 1 | 0 | 0 |
| $E' \, \Gamma_6$ | 2 | 2 | -1 | -1 | 2 | -1 | -1 | 0 | 0 |
| $\bar{E}_1 \, \Gamma_7$ | 2 | -2 | 1 | -1 | 0 | $\sqrt{3}$ | $-\sqrt{3}$ | 0 | 0 |
| $\bar{E}_2 \, \Gamma_8$ | 2 | -2 | 1 | -1 | 0 | $-\sqrt{3}$ | $\sqrt{3}$ | 0 | 0 |
| $\bar{E}_3 \, \Gamma_9$ | 2 | -2 | -2 | 2 | 0 | 0 | 0 | 0 | 0 |

**Table S1.** Character table at the zone center ($\Gamma$ point) of monolayer WSe$_2$ with the $D_{3h}$ point double group. We use $\omega = e^{i\frac{2\pi}{3}}$.

| $C_{3h}$ | $E$ | $C_3^+$ | $C_3^-$ | $\sigma_h$ | $S_3^+$ | $S_3^-$ | $\bar{E}$ | $\bar{C}_3^+$ | $\bar{C}_3^-$ | $\bar{\sigma}_h$ | $\bar{S}_3^+$ | $\bar{S}_3^-$ |
|---|---|---|---|---|---|---|---|---|---|---|---|---|
| $A'$  $K_1$ | 1 | 1 | 1 | 1 | 1 | 1 | 1 | 1 | 1 | 1 | 1 | 1 |
| $^2E'$  $K_2$ | 1 | $\omega^*$ | $\omega$ | 1 | $\omega^*$ | $\omega$ | 1 | $\omega^*$ | $\omega$ | 1 | $\omega^*$ | $\omega$ |
| $^1E'$  $K_3$ | 1 | $\omega$ | $\omega^*$ | 1 | $\omega$ | $\omega^*$ | 1 | $\omega$ | $\omega^*$ | 1 | $\omega$ | $\omega^*$ |
| $A''$  $K_4$ | 1 | 1 | 1 | $-1$ | $-1$ | $-1$ | 1 | 1 | 1 | $-1$ | $-1$ | $-1$ |
| $^2E''$  $K_5$ | 1 | $\omega^*$ | $\omega$ | $-1$ | $-\omega^*$ | $-\omega$ | 1 | $\omega^*$ | $\omega$ | $-1$ | $-\omega^*$ | $-\omega$ |
| $^1E''$  $K_6$ | 1 | $\omega$ | $\omega^*$ | $-1$ | $-\omega$ | $-\omega^*$ | 1 | $\omega$ | $\omega^*$ | $-1$ | $-\omega$ | $-\omega^*$ |
| $^1\bar{E}_3$  $K_7$ | 1 | $-\omega^*$ | $-\omega$ | $-i$ | $i\omega^*$ | $-i\omega$ | $-1$ | $\omega^*$ | $\omega$ | $i$ | $-i\omega^*$ | $i\omega$ |
| $^2\bar{E}_3$  $K_8$ | 1 | $-\omega$ | $-\omega^*$ | $i$ | $-i\omega$ | $i\omega^*$ | $-1$ | $\omega$ | $\omega^*$ | $-i$ | $i\omega$ | $-i\omega^*$ |
| $^2\bar{E}_2$  $K_9$ | 1 | $-\omega^*$ | $-\omega$ | $i$ | $-i\omega^*$ | $i\omega$ | $-1$ | $\omega^*$ | $\omega$ | $-i$ | $i\omega^*$ | $-i\omega$ |
| $^1\bar{E}_2$  $K_{10}$ | 1 | $-\omega$ | $-\omega^*$ | $-i$ | $i\omega$ | $-i\omega^*$ | $-1$ | $\omega$ | $\omega^*$ | $i$ | $-i\omega$ | $i\omega^*$ |
| $^1\bar{E}_1$  $K_{11}$ | 1 | $-1$ | $-1$ | $-i$ | $i$ | $-i$ | $-1$ | 1 | 1 | $i$ | $-i$ | $i$ |
| $^2\bar{E}_1$  $K_{12}$ | 1 | $-1$ | $-1$ | $i$ | $-i$ | $i$ | $-1$ | 1 | 1 | $-i$ | $i$ | $-i$ |

**Table S2.** Character table at the zone corners ($K$ and $K'$ points) of monolayer WSe$_2$ with the $C_{3h}$ point double group. We use $\omega = e^{i\frac{2\pi}{3}}$.

In the main paper, we have used the group theory to derive the selection rules of intervalley transitions. Monolayer WSe$_2$ crystal possesses the $D_{3h}$ point group. The electronic and phonon states at the zone center ($\Gamma$ point) have the $D_{3h}$ double point group (including the spin). The electronic and phonon states at the zone corner (K and K' points) have the $C_{3h}$ double point group (including the spin). Table S1 and S2 display the irreducible representations and characters for the zone-center and zone-corner states in monolayer WSe$_2$ [13]. Both tables are adopted from Ref. [13]. But the characters here are taken as the complex conjugate of those in Ref. [13] in order to stay consistent with the convention used in our previous paper [1]. The irreducible representations of relavant electronic and phonon states are denoted in Fig. 1 of the main paper.

We define the chirality of a phonon by the character of the operator $C_3^+$, which rotates the wavefunction counter-clockwise for 120 degree (or equivalently, rotates the coordinate clockwise for 120 degree). In our convention, the right-handed chiral phonon ($\Omega^+$) transforms as $C_3^+ | \Omega^+ \rangle = \omega^* | \Omega^+ \rangle$, where $\omega = e^{i\frac{2\pi}{3}}$. The left-handed chiral phonon ($\Omega^-$) transforms as $C_3^+ | \Omega^- \rangle = \omega | \Omega^- \rangle$. The phonon with K$_3$ representation is thus left-handed chiral phonon ($\Omega^-$), as denoted in Fig. 1 of the main paper.

## 9. First-principles calculations of exciton exchange interaction

We have carried out first-principles calculations to obtain the splitting energy between the intravalley dark exciton ($D^0$) and intervalley exciton ($I^0$). The binding energy of an exciton is contributed by the direct electron-hole Coulomb interaction and electron-hole exchange interaction. $D^0$ and $I^0$ have the same effective mass, so their direct Coulomb interaction is the same. But $D^0$ with antiparallel electron spins has no exchange interaction, whereas $I^0$ with parallel electron spins has finite exchange interaction. So, the energy separation between $D^0$ and $I^0$ is caused by the exchange interaction of $I^0$.

In our theoretical analysis, we describe different exciton states as:

Intravalley dark exciton: $\mid D^0 \rangle = \sum_{\mathbf{k}} \varphi_D(\mathbf{k}) \mid v_{\uparrow,\mathbf{k}}^h; c_{\downarrow,\mathbf{k}} \rangle$, (S2)

Intravalley bright exciton: $\mid A^0 \rangle = \sum_{\mathbf{k}} \varphi_A(\mathbf{k}) \mid v_{\uparrow,\mathbf{k}}^h; c_{\uparrow,\mathbf{k}} \rangle$, (S3)

Intervalley dark exciton: $\mid I^0 \rangle = \sum_{\mathbf{k}} \varphi_I(\mathbf{k}) \mid v_{\uparrow,\mathbf{k}}^h; \tilde{c}_{\uparrow,\mathbf{k}} \rangle$, (S4)

Here $\mid v_{\uparrow,\mathbf{k}}^h \rangle$ denotes the Bloch state of a hole in the highest valence band in the K valley. $\mid c_{\downarrow,\mathbf{k}} \rangle$ denotes the Bloch state of an electron in the lower conduction band in the K valley. $\mid c_{\uparrow,\mathbf{k}} \rangle$ denotes the Bloch state of an electron in the upper conduction band in the K valley. $\mid \tilde{c}_{\uparrow,\mathbf{k}} \rangle$ denotes the Bloch state of an electron in the lower conduction band in the K' valley.

The matrix elements of the electron-hole interaction for the excitons defined in Eq. (S2 – S4) are given, respectively, by:

$\langle v_{\uparrow,\mathbf{k}}^h; c_{\downarrow,\mathbf{k}} \mid V \mid v_{\uparrow,\mathbf{k}'}^h; c_{\downarrow,\mathbf{k}'} \rangle = - \langle v_{\uparrow,\mathbf{k}}^h; c_{\downarrow,\mathbf{k}} \mid W \mid v_{\uparrow,\mathbf{k}'}^h; c_{\downarrow,\mathbf{k}'} \rangle$ (S5)

$\langle v_{\uparrow,\mathbf{k}}^h; c_{\uparrow,\mathbf{k}} \mid V \mid v_{\uparrow,\mathbf{k}'}^h; c_{\uparrow,\mathbf{k}'} \rangle = - \langle v_{\uparrow,\mathbf{k}}^h; c_{\uparrow,\mathbf{k}} \mid W \mid v_{\uparrow,\mathbf{k}'}^h; c_{\uparrow,\mathbf{k}'} \rangle + \langle v_{\uparrow,\mathbf{k}}^h; c_{\uparrow,\mathbf{k}} \mid V_{ex} \mid v_{\uparrow,\mathbf{k}'}^h; c_{\uparrow,\mathbf{k}'} \rangle$ (S6)

$\langle v_{\uparrow,\mathbf{k}}^h; \tilde{c}_{\uparrow,\mathbf{k}} \mid V \mid v_{\uparrow,\mathbf{k}'}^h; \tilde{c}_{\uparrow,\mathbf{k}'} \rangle = - \langle v_{\uparrow,\mathbf{k}}^h; \tilde{c}_{\uparrow,\mathbf{k}} \mid W \mid v_{\uparrow,\mathbf{k}'}^h; \tilde{c}_{\uparrow,\mathbf{k}'} \rangle + \langle v_{\uparrow,\mathbf{k}}^h; \tilde{c}_{\uparrow,\mathbf{k}} \mid V_{ex} \mid v_{\uparrow,\mathbf{k}'}^h; \tilde{c}_{\uparrow,\mathbf{k}'} \rangle$ (S7)

Here the W terms denote the direct integral for screened Coulomb interaction; The $V_{ex}$ terms denote the electron-hole exchange for the bare Coulomb interaction. We have [14]:

$$\langle v_{\uparrow,\mathbf{k}}^h; \tilde{c}_{\uparrow,\mathbf{k}} \mid W \mid v_{\uparrow,\mathbf{k}'}^h; \tilde{c}_{\uparrow,\mathbf{k}'} \rangle = \frac{1}{\Omega} \sum_{\mathbf{G},\mathbf{G}'} W_{\mathbf{G},\mathbf{G}'}(\mathbf{q}) \langle v_{\uparrow,\mathbf{k}'}^h \mid e^{i(\mathbf{G}+\mathbf{q})\cdot\mathbf{r}} \mid v_{\uparrow,\mathbf{k}}^h \rangle \langle \tilde{c}_{\uparrow,\mathbf{k}} \mid e^{-i(\mathbf{G}+\mathbf{q})\cdot\mathbf{r}} \mid \tilde{c}_{\uparrow,\mathbf{k}'} \rangle$$
(S8)

$$\langle v_{\uparrow,\mathbf{k}}^h; \tilde{c}_{\uparrow,\mathbf{k}} \mid V_{ex} \mid v_{\uparrow,\mathbf{k}'}^h; \tilde{c}_{\uparrow,\mathbf{k}'} \rangle = \frac{1}{\Omega} \sum_{\mathbf{G}} \frac{4\pi e^2}{|\mathbf{G}+\mathbf{Q}|^2} \langle \tilde{c}_{\uparrow,\mathbf{k}} \mid e^{i(\mathbf{G}+\mathbf{Q})\cdot\mathbf{r}} \mid v_{\uparrow,\mathbf{k}}^h \rangle \langle v_{\uparrow,\mathbf{k}'}^h \mid e^{-i(\mathbf{G}+\mathbf{Q})\cdot\mathbf{r}} \mid \tilde{c}_{\uparrow,\mathbf{k}'} \rangle.$$ (S9)

Here $\Omega$ is the volume of the sample; $\mathbf{G}$'s denotes the reciprocal lattice vectors of the 3D superlattice; $\mathbf{q} = \mathbf{k}' - \mathbf{k}$ and $\mathbf{Q}$ are the exciton momentum. For intravalley exciton, we take $\mathbf{Q} = 0$ (neglecting the photon momentum). For intervalley exciton, we take $\mathbf{Q} = \mathbf{K}' - \mathbf{K}$. The Bloch states are obtained by LASTO code via a supercell method. So, the underlying basis functions are 3D augmented plane waves.

Since the electron-hole exchange term ($V_{ex}$) is much weaker than the direct term ($W$), we can treat it as a perturbation. We first neglect the $V_{ex}$ term and solve the quasi-2D exciton problem via the envelope-function method. In this method, the kinetic term is replaced by an effective mass Hamiltonian and the potential term is replaced by a Keldysh potential. Such a model potential has been shown to give adequate description of the low-lying exciton Rydberg states (up to $n = 5$) in monolayer WSe$_2$, including their energy levels and diamagnetic shifts [15].

Since the $W$ terms are practically the same for the intravalley and intervalley excitons, the exciton envelope function obtained by omitting $V_x$ can be described by $\varphi_D(\mathbf{k})$ in Eq. (S2). We may assume that the exciton envelope functions are the same for both intravalley and intervalley excitons. We then evaluate the correction to the exciton binding energy for $A^0$ exciton and $I^0$ exciton via the first-order perturbation theory. The correction energy for the intervalley $I^0$ exciton is:

$$\Delta E_X = \frac{L}{\Omega L_c} \sum_{\mathbf{k}} \varphi_D^*(\mathbf{k}) \sum_{\mathbf{k}'} \varphi_D(\mathbf{k}') \left\langle v_{\uparrow,\mathbf{k}}^h; \tilde{c}_{\uparrow,\mathbf{k}} | V_x | v_{\uparrow,\mathbf{k}'}^h; \tilde{c}_{\uparrow,\mathbf{k}'} \right\rangle$$

$$= \frac{1}{AL_c} \sum_{\mathbf{G}} \frac{4\pi e^2}{|\mathbf{G}+\mathbf{Q}|^2} \sum_{\mathbf{k}} \varphi_D^*(\mathbf{k}) \left\langle \tilde{c}_{\uparrow,\mathbf{k}} | e^{i(\mathbf{G}+\mathbf{Q})\cdot\mathbf{r}} | v_{\uparrow,\mathbf{k}}^h \right\rangle \sum_{\mathbf{k}'} \varphi_D(\mathbf{k}') \left\langle v_{\uparrow,\mathbf{k}'}^h | e^{-i(\mathbf{G}+\mathbf{Q})\cdot\mathbf{r}} | \tilde{c}_{\uparrow,\mathbf{k}'} \right\rangle$$

$$= \frac{1}{L_c} \sum_{\mathbf{G}} \frac{4\pi e^2}{|\mathbf{G}+\mathbf{Q}|^2} \left| \int \frac{d\mathbf{k}}{(2\pi)^2} \varphi_D^*(\mathbf{k}) \left\langle \tilde{c}_{\uparrow,\mathbf{k}} | e^{i(\mathbf{G}+\mathbf{Q})\cdot\mathbf{r}} | v_{\uparrow,\mathbf{k}}^h \right\rangle \right|^2$$

$$= \left| \int \frac{d\mathbf{k}}{(2\pi)^2} \varphi_D^*(\mathbf{k}) D(\mathbf{Q},\mathbf{k}) \right|^2$$

$$\approx D(\mathbf{Q},\mathbf{0}) \left| \int \frac{d\mathbf{k}}{(2\pi)^2} \varphi_D^*(\mathbf{k}) \right|^2. \tag{S10}$$

Here $D(\mathbf{Q},\mathbf{k}) = \frac{1}{L_c} \sum_{\mathbf{G}} \frac{4\pi e^2}{|\mathbf{G}+\mathbf{Q}|^2} \left| \left\langle \tilde{c}_{\uparrow,\mathbf{k}} | e^{i(\mathbf{G}+\mathbf{Q})\cdot\mathbf{r}} | v_{\uparrow,\mathbf{k}}^h \right\rangle \right|^2$; $L$ ($L_c$) denotes the length of the sample (supercell) along the out-of-plane axis; $\mathbf{k}$ denotes the wave-vector within the 2D Brillouin zone. The envelope function $\varphi_D(\mathbf{k})$ is normalized according to $\sum_{\mathbf{k}} |\varphi_D(\mathbf{k})|^2 = 1$.

In the augmented plane wave (APW) method, the calculation of matrix elements $\left\langle \tilde{c}_{\uparrow,\mathbf{k}} | e^{i(\mathbf{G}+\mathbf{Q})\cdot\mathbf{r}} | v_{\uparrow,\mathbf{k}}^h \right\rangle$ contains integrals for the muffin-tin sphere part and the interstitial part [16]. Those integrals have been derived in Ref. [16]. We implemented those integrals in the LASTO (Linear augmented Slater-type orbital) package [17, 18] and used them to calculate the electron-hole exchange interaction $D(\mathbf{Q},\mathbf{k})$. The LASTO package can produce results similar to those of Wien2k [19, 20]. We have included the spin-orbit interaction in the calculation. We obtain $D(\mathbf{Q}) = 4.17$ eV Å$^2$ for the intravalley $A^0$ exciton ($\mathbf{Q} = 0$) and $D(\mathbf{Q}) = 4.80$ eV Å$^2$ for the intervalley $I^0$ exciton ($\mathbf{Q} = \mathbf{K}' - \mathbf{K}$).

To model the dark exciton $D^0$, we adopt an electron effective mass of $0.46m_0$ based on our result of trion splitting in the main paper ($m_0$ is the free electron mass). The hole effective mass is taken as $0.42m_0$ based on the average of published values (which range from $0.36m_0$ to $0.51m_0$ [21-23]) and the ARPES result ($0.42m_0$) [9]. This leads to a reduced mass of $0.219m_0$ for the dark exciton $D^0$. For comparison, the reduced mass for the bright exciton was established to be ~$0.2m_0$ [24]. Thus, the binding energy of $D^0$ should be ~9% larger than that of $A^0$ exciton. Using the same model potential as in Ref. [15], we can calculate the $D^0$ exciton envelope function. We get:

$$F_D(0) \equiv \int \frac{d\mathbf{k}}{(2\pi)^2} \varphi_D^*(\mathbf{k}) = 0.464/a_X, \tag{S10}$$

where $a_X \approx 9.59$ Å is the effective exciton Bohr radius. Finally, we obtain the correction energy due to the exchange interaction to be $\Delta E_X \approx 11.2$ meV for the intervalley $I^0$ exciton.

The value is consistent with the result (~11 meV) of another theoretical study [25]. The result is close to our observed energy separation (10 meV) between the $D^0$ and $I^0$ luminescence lines in experiment.

The small deviation between theory and experiment might be due to the renormalization of electron mass. Our measured electron mass ($0.46m_0$) is obtained from the hole-side trion data, while our calculation here requires the mass at charge neutrality. A prior theoretical study [26] suggests that the electron effective mass in trion can be ~17% heavier than the mass at charge neutrality due to the mass renormalization by the trion-lattice coupling. By considering this effect, we may set the charge-neutrality electron effective mass to be in the range of $0.4m_0$ to $0.46m_0$. The corresponding range of correction energy for the intervalley $I^0$ exciton is $\Delta E_X \approx 9.8 - 11.2$ meV. The range matches excellently our observed energy separation (10 meV) between the $D^0$ and $I^0$ lines. Our calculation therefore strongly supports that the $I^0$ line in experiment comes from the intervalley exciton.

Similarly, we have calculated the correction energy due to the exchange interaction for the intravalley $A^0$ exciton. The result is $\Delta E_X \approx 7.9$ meV, which is ~3 meV smaller than the correction energy of the $I^0$ exciton. In experiment, the separation between the $A^0$ and $I^0$ lines is 31 meV. If we assume that the $A^0$ and $I^0$ exciton have the same effective mass, then the separation between the upper and lower conduction bands should be ~34 meV. But in reality, the $I^0$ exciton has slightly larger effective mass than $A^0$. So, the separation between the two conduction bands should be smaller than 34 meV.